\documentclass[10pt, a4paper, conference]{IEEEtran}
\usepackage{url}
\usepackage[utf8]{inputenc}
\usepackage{xcolor}
\usepackage{amsmath}
\usepackage{multirow}
\usepackage{amssymb}
\usepackage{enumitem}

\usepackage{gensymb}

\usepackage[acronyms,nonumberlist,nopostdot,nomain,nogroupskip]{glossaries}
\usepackage{tablefootnote}
\usepackage{booktabs}
\usepackage{tabularx}
\usepackage{epsfig}
\usepackage[outdir=img/]{epstopdf}
\usepackage{tikz}
\usepackage{pgfplots}
\pgfplotsset{compat=newest} 
\pgfplotsset{plot coordinates/math parser=false} 
\newlength\fheight
\newlength\fwidth
\usetikzlibrary{plotmarks,patterns,decorations.pathreplacing,backgrounds,calc,arrows,arrows.meta,spy,matrix}
\usepgfplotslibrary{patchplots,groupplots}
\usepackage{tikzscale}
\usepackage{siunitx}

\usepackage{multirow}
\usepackage{tkz-kiviat}

\usepackage{multirow}
\usepackage{tkz-kiviat}

\usepackage[font=scriptsize]{subcaption}
\usepackage[font=footnotesize]{caption}

\usepackage{mathtools}

\usepackage{dblfloatfix}    
\usepackage{colortbl}

\newacronym{3gpp}{3GPP}{3rd Generation Partnership Project}
\newacronym{adc}{ADC}{Analog to Digital Converter}
\newacronym{5g}{5G}{5th generation}
\newacronym{aimd}{AIMD}{Additive Increase Multiplicative Decrease}
\newacronym{am}{AM}{Acknowledged Mode}
\newacronym{amc}{AMC}{Adaptive Modulation and Coding}
\newacronym{ap}{AP}{Average Precision}
\newacronym{aqm}{AQM}{Active Queue Management}
\newacronym{awgn}{AGWN}{Additive White Gaussian Noise}
\newacronym{balia}{BALIA}{Balanced Link Adaptation}
\newacronym{bdp}{BDP}{Bandwidth-Delay Product}
\newacronym{bf}{BF}{beamforming}
\newacronym{cc}{CC}{Congestion Control}
\newacronym{cdf}{CDF}{Cumulative Distribution Function}
\newacronym{cn}{CN}{Core Network}
\newacronym{cqi}{CQI}{Channel Quality Information}
\newacronym{cp}{CP}{Control Plane}
\newacronym{csirs}{CSI-RS}{Channel State Information - Reference Signal}
\newacronym{dc}{DC}{Dual Connectivity}
\newacronym{rb}{RB}{Resource Block}
\newacronym{dce}{DCE}{Direct Code Execution}
\newacronym{dci}{DCI}{Downlink Control Information}
\newacronym{udp}{UDP}{User Datagram Protocol}
\newacronym{dl}{DL}{Downlink}
\newacronym{dmr}{DMR}{Deadline Miss Ratio}
\newacronym{dmrs}{DMRS}{DeModulation Reference Signal}
\newacronym{e2e}{E2E}{End-to-End}
\newacronym{ppp}{PPP}{Poission Point Process}
\newacronym{si}{SI}{Study Item}
\newacronym{ecn}{ECN}{Explicit Congestion Notification}
\newacronym{edf}{EDF}{Earliest Deadline First}
\newacronym{enb}{eNB}{eNodeB}
\newacronym{epc}{EPC}{Evolved Packet Core}
\newacronym{es}{ES}{Edge Server}
\newacronym{cav}{CAV}{Connected and Autonomous Vehicle}
\newacronym{fdma}{FDMA}{Frequency Division Multiple Access}
\newacronym{fdd}{FDD}{Frequency Division Duplexing}
\newacronym{upa}{UPA}{Uniform Planar Array}
\newacronym[firstplural=Radio Access Technologies (RATs)]{rat}{RAT}{Radio Access Technology}
\newacronym[firstplural=Radio Access Technology (RTs)]{rt}{RT}{Radio Technology}
\newacronym{fs}{FS}{Fast Switching}
\newacronym{isd}{ISD}{inter-site distance}
\newacronym{ftp}{FTP}{File Transfer Protocol}
\newacronym{gnb}{gNB}{Next Generation Node Base}
\newacronym{harq}{HARQ}{Hybrid Automatic Repeat reQuest}
\newacronym{hetnet}{HetNet}{Heterogeneous Network}
\newacronym{hh}{HH}{Hard Handover}
\newacronym{hol}{HOL}{Head-of-Line}
\newacronym{ia}{IA}{Initial Access}
\newacronym{imt}{IMT}{International Mobile Telecommunication}
\newacronym{iot}{IoT}{Internet of Things}
\newacronym{los}{LOS}{Line of Sight}
\newacronym{lte}{LTE}{Long Term Evolution}
\newacronym{m2m}{M2M}{Machine to Machine}
\newacronym{mac}{MAC}{Medium Access Control}
\newacronym{mc}{MC}{Multi-Connectivity}
\newacronym{mcs}{MCS}{Modulation and Coding Scheme}
\newacronym{mec}{MEC}{Mobile Edge Cloud}
\newacronym{mi}{MI}{Mutual Information}
\newacronym{mimo}{MIMO}{Multiple Input Multiple Output}
\newacronym{mmwave}{mmWave}{millimeter wave}
\newacronym{mptcp}{MPTCP}{Multipath TCP}
\newacronym{mr}{MR}{Maximum Rate}
\newacronym{mss}{MSS}{Maximum Segment Size}
\newacronym{mtd}{MTD}{Machine-Type Device}
\newacronym{mtu}{MTU}{Maximum Transmission Unit}
\newacronym{nfv}{NFV}{Network Function Virtualization}
\newacronym{nlos}{NLOS}{Non Line of Sight}
\newacronym{nlosb}{NLOSb}{Building Non Line of Sight}
\newacronym{nlosv}{NLOSv}{Vehicle Non Line of Sight}
\newacronym{nr}{NR}{New Radio}
\newacronym{ofdm}{OFDM}{Orthogonal Frequency Division Multiplexing}
\newacronym{pdcch}{PDCCH}{Physical Downlonk Control Channel}
\newacronym{pdcp}{PDCP}{Packet Data Convergence Protocol}
\newacronym{pdsch}{PDSCH}{Physical Downlink Shared Channel}
\newacronym{pdu}{PDU}{Packet Data Unit}
\newacronym{pf}{PF}{Proportional Fair}
\newacronym{pgw}{PGW}{Packet Gateway}
\newacronym{phy}{PHY}{Physical}
\newacronym{pbch}{PBCH}{Physical Broadcast Channel}
\newacronym[plural=\gls{mme}s,firstplural=Mobility Management Entities (MMEs)]{mme}{MME}{Mobility Management Entity}
\newacronym{prb}{PRB}{Physical Resource Block}
\newacronym{pss}{PSS}{Primary Synchronization Signal}
\newacronym{pucch}{PUCCH}{Physical Uplink Control Channel}
\newacronym{pusch}{PUSCH}{Physical Uplink Shared Channel}
\newacronym{rach}{RACH}{Random Access Channel}
\newacronym{ran}{RAN}{Radio Access Network}
\newacronym{red}{RED}{Random Early Detection}
\newacronym{rf}{RF}{Radio Frequency}
\newacronym{rlc}{RLC}{Radio Link Control}
\newacronym{rlf}{RLF}{Radio Link Failure}
\newacronym{rrc}{RRC}{Radio Resource Control}
\newacronym{rrm}{RRM}{Radio Resource Management}
\newacronym{rr}{RR}{Round Robin}
\newacronym{rs}{RS}{Remote Server}
\newacronym{rsrp}{RSRP}{Reference Signal Received Power}
\newacronym{rss}{RSS}{Received Signal Strength}
\newacronym{rtt}{RTT}{Round Trip Time}
\newacronym{rw}{RW}{Receive Window}
\newacronym{rx}{RX}{Receiver}
\newacronym{sa}{SA}{standalone}
\newacronym{sack}{SACK}{Selective Acknowledgment}
\newacronym{sap}{SAP}{Service Access Point}
\newacronym{sch}{SCH}{Secondary Cell Handover}
\newacronym{scoot}{SCOOT}{Split Cycle Offset Optimization Technique}
\newacronym{sdma}{SDMA}{Spatial Division Multiple Access}
\newacronym{sinr}{SINR}{Signal to Interference plus Noise Ratio}
\newacronym{sm}{SM}{Saturation Mode}
\newacronym{snr}{SNR}{Signal to Noise Ratio}
\newacronym{son}{SON}{Self-Organizing Network}
\newacronym{ss}{SS}{Synchronization Signal}
\newacronym{srs}{SRS}{Sounding Reference Signal}
\newacronym{sss}{SSS}{Secondary Synchronization Signal}
\newacronym{tb}{TB}{Transport Block}
\newacronym{tcp}{TCP}{Transmission Control Protocol}
\newacronym{tdd}{TDD}{Time Division Duplexing}
\newacronym{tdma}{TDMA}{Time Division Multiple Access}
\newacronym{tfl}{TfL}{Transport for London}
\newacronym{tm}{TM}{Transparent Mode}
\newacronym{prr}{PRR}{Packet Reception Ratio}
\newacronym{trp}{TRP}{Transmitter Receiver Pair}
\newacronym{tti}{TTI}{Transmission Time Interval}
\newacronym{ttt}{TTT}{Time-to-Trigger}
\newacronym{tx}{TX}{Transmitter}
\newacronym{ue}{UE}{User Equipment}
\newacronym{ul}{UL}{Uplink}
\newacronym{uml}{UML}{Unified Modeling Language}
\newacronym{um}{UM}{Unacknowledged Mode}
\newacronym{utc}{UTC}{Urban Traffic Control}
\newacronym{vm}{VM}{Virtual Machine}
\newacronym{rsrq}{RSRQ}{Reference Signal Received Quality}
\newacronym{rssi}{RSSI}{Received Signal Strength Indicator}
\newacronym{crs}{CRS}{Cell Reference Signal}
\newacronym{v2v}{V2V}{Vehicle-to-Vehicle}
\newacronym{v2i}{V2I}{Vehicle-to-Infrastructure}
\newacronym{v2n}{V2N}{Vehicle-to-Network}
\newacronym{v2x}{V2X}{Vehicle-to-Everything}
\newacronym{vn}{VN}{Vehicular Node}
\newacronym{dsrc}{DSRC}{Dedicated Short Range Communication}
\newacronym{ci}{CI}{context information}
\newacronym{voi}{VoI}{value of information}
\newacronym{gps}{GPS}{Global Positioning System}
\newacronym{qos}{QoS}{Quality of Service}
\newacronym{qoe}{QoE}{Quality of Experience}
\newacronym{ml}{ML}{Machine Learning}
\newacronym{ahp}{AHP}{Analytic Hierarchy Process}
\newacronym{lidar}{LiDAR}{Light Detection and Ranging}
\newacronym{sumo}{SUMO}{Simulation of Urban MObility}
\newacronym{wave}{WAVE}{Wireless Access in Vehicular Environment}
\newacronym{c-its}{C-ITS}{Connected Intelligent Transportation System}
\newacronym{dash}{DASH}{Dynamic Adaptive Streaming over HTTP}
\newacronym{http}{HTTP}{HyperText Transfer Protocol}
\newacronym{dcnn}{DCNN}{Deep Convolutional Neural Network}
\newacronym{cnn}{CNN}{Convolutional Neural Network}
\newacronym{mdp}{MDP}{Markov Decision Process}
\newacronym{rl}{RL}{Reinforcement Learning}
\newacronym{drl}{DRL}{Deep Reinforcement Learning}
\newacronym{map}{mAP}{mean Average Precision}
\newacronym{avod}{AVOD}{Aggregate View Object Detection}
\newacronym{yolo}{YOLO}{You Only Look Once}
\newacronym{bev}{BEV}{bird's eye view}
\newacronym{rpn}{RPN}{Region Proposal Network}
\newacronym{iou}{IoU}{Intersection over Union}
\newacronym{roi}{RoI}{Region of Interest}
\newacronym{nms}{NMS}{non-maximum suppression}
\newacronym{vfe}{VFE}{Voxel Feature Encoding}
\newacronym{fov}{FoV}{field of view}
\tikzstyle{startstop} = [rectangle, rounded corners, minimum width=2cm, minimum height=0.5cm,text centered, draw=black]
\tikzstyle{io} = [trapezium, trapezium left angle=70, trapezium right angle=110, minimum width=3cm, minimum height=1cm, text centered, draw=black]
\tikzstyle{process} = [rectangle, minimum width=2cm, minimum height=0.5cm, text centered, draw=black, alignb=center]
\tikzstyle{decision} = [ellipse, minimum width=2cm, minimum height=1cm, text centered, draw=black]
\tikzstyle{arrow} = [thick,<->,>=stealth]
\tikzstyle{line} = [thick,>=stealth]
\tikzstyle{darrow} = [thick,<->,>=stealth,dashed]
\tikzstyle{sarrow} = [thick,->,>=stealth]
\tikzstyle{larrow} = [line width=0.1mm,dashdotted,->,>=stealth]

\makeatletter
\def\grd@save@target#1{%
  \def\grd@target{#1}}
\def\grd@save@start#1{%
  \def\grd@start{#1}}
\tikzset{
  grid with coordinates/.style={
    to path={%
      \pgfextra{%
        \edef\grd@@target{(\tikztotarget)}%
        \tikz@scan@one@point\grd@save@target\grd@@target\relax
        \edef\grd@@start{(\tikztostart)}%
        \tikz@scan@one@point\grd@save@start\grd@@start\relax
        \draw[minor help lines] (\tikztostart) grid (\tikztotarget);
        \draw[major help lines] (\tikztostart) grid (\tikztotarget);
        \grd@start
        \pgfmathsetmacro{\grd@xa}{\the\pgf@x/1cm}
        \pgfmathsetmacro{\grd@ya}{\the\pgf@y/1cm}
        \grd@target
        \pgfmathsetmacro{\grd@xb}{\the\pgf@x/1cm}
        \pgfmathsetmacro{\grd@yb}{\the\pgf@y/1cm}
        \pgfmathsetmacro{\grd@xc}{\grd@xa + \pgfkeysvalueof{/tikz/grid with coordinates/major step x}}
        \pgfmathsetmacro{\grd@yc}{\grd@ya + \pgfkeysvalueof{/tikz/grid with coordinates/major step y}}
        \foreach \x in {\grd@xa,\grd@xc,...,\grd@xb}
        \node[anchor=north] at (\x,\grd@ya) {\pgfmathprintnumber{\x}};
        \foreach \y in {\grd@ya,\grd@yc,...,\grd@yb}
        \node[anchor=east] at (\grd@xa,\y) {\pgfmathprintnumber{\y}};
      }
    }
  },
  minor help lines/.style={
    help lines,
    gray,
    line cap =round,
    xstep=\pgfkeysvalueof{/tikz/grid with coordinates/minor step x},
    ystep=\pgfkeysvalueof{/tikz/grid with coordinates/minor step y}
  },
  major help lines/.style={
    help lines,
    line cap =round,
    line width=\pgfkeysvalueof{/tikz/grid with coordinates/major line width},
    xstep=\pgfkeysvalueof{/tikz/grid with coordinates/major step x},
    ystep=\pgfkeysvalueof{/tikz/grid with coordinates/major step y}
  },
  grid with coordinates/.cd,
  minor step x/.initial=.5,
  minor step y/.initial=.2,
  major step x/.initial=1,
  major step y/.initial=1,
  major line width/.initial=1pt,
}
\makeatother

\usepackage[capitalize]{cleveref}
\definecolor{steelblue}{RGB}{176,196,222}

\makeglossaries

\begin{document}


\title{On the Role of Sensor Fusion for Object Detection in Future Vehicular Networks}

\author{\IEEEauthorblockN{Valentina Rossi,
                          Paolo Testolina,
                          Marco Giordani,
                          Michele Zorzi}
        \IEEEauthorblockA{\vspace{0.2cm}     Department of Information Engineering, University of Padova, Italy \\ Email: \texttt{\{rossivalen, testolina, giordani, zorzi\}@dei.unipd.it}},
        \vspace{-5ex}%
}

\maketitle

\glsunset{nr}

\begin{abstract}
Fully autonomous driving systems require fast detection and recognition of sensitive objects in the environment.
In this context, intelligent vehicles should share their sensor data with computing platforms and/or other vehicles, to detect objects beyond their own sensors' fields of view. However, the resulting huge volumes of data to be exchanged can be challenging to handle for standard communication technologies.
In this paper, we evaluate how using a combination of different sensors affects the detection of the environment in which the vehicles move and operate. The final objective is to identify the optimal setup that would minimize the amount of data to be distributed over the channel, with negligible degradation in terms of object detection accuracy.
To this aim, we extend an already available object detection algorithm so that it can consider, as an input, camera images, LiDAR point clouds, or a combination of the two, and compare the accuracy performance of the different approaches using two realistic datasets. 
Our results show that, although sensor fusion always achieves more accurate detections, LiDAR only inputs can obtain similar results for large objects while mitigating the burden on the channel.
\end{abstract}
\begin{IEEEkeywords}
Autonomous driving, object detection, sensor fusion, machine learning, AVOD.
\end{IEEEkeywords}
		\begin{tikzpicture}[remember picture,overlay]
\node[anchor=north,yshift=-10pt] at (current page.north) {\parbox{\dimexpr\textwidth-\fboxsep-\fboxrule\relax}{
\centering\footnotesize This paper has been accepted for presentation at the Joint European Conference on Networks and Communications \& 6G Summit, \textcopyright 2021 IEEE.\\
Please cite it as: V. Rossi, P. Testolina, M. Giordani, M. Zorzi, “On the Role of Sensor Fusion for Object Detection in Future Vehicular Networks”, \\ Joint European Conference on Networks and Communications \& 6G Summit (EuCNC/6G Summit), 2021.}};
\end{tikzpicture}

\section{Introduction}
\label{sec:intro}
The worldwide commercialization of \gls{5g} wireless networks is pushing towards the development of \glspl{cav} to improve road safety and decrease traffic congestion and carbon emissions~\cite{clements2017economic}.
For these systems to be truly autonomous, intelligent vehicles must be able to both identify and track critical entities in the environment (e.g., cars, pedestrians, cyclists, etc.), a task that is typically referred to as \emph{object detection}.

To this aim, autonomous vehicles are usually equipped with different sensors, each of which has its own strengths and weaknesses.
On one hand, \gls{lidar} sensors provide both object detection and classification  in all lighting conditions and are currently the most precise sensors to measure range. However, they generate point clouds that are quite difficult to compress with low latency~\cite{varischio2021hybrid}, are sensitive to diffraction effects (especially in closed environments like tunnels), and offer a sparse representation of the surroundings, particularly when considering faraway objects.
On the other hand, RGB cameras are much cheaper than \glspl{lidar}, and less of a computational burden for both compression and data exchange. In turn, they suffer from strong sensitivity to external illumination and visibility conditions and have a limited \gls{fov}, and therefore cannot provide detailed depth information~\cite{secci2020failures}.

\begin{figure*}[t]
    \centering
    \includegraphics[width=0.99\textwidth]{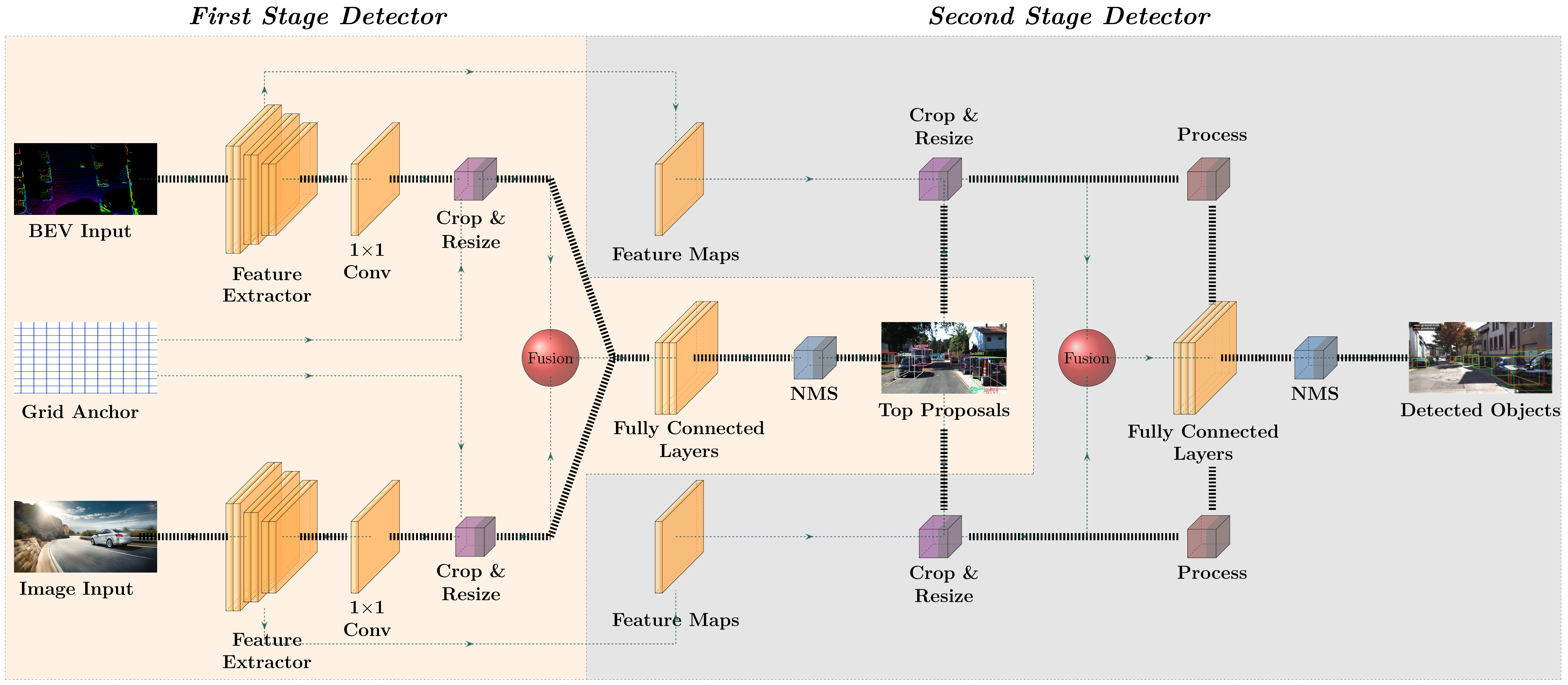}
    \caption{The AVOD architecture. The first stage detector consists of an RPN to regress non-oriented anchor proposals, while the second stage detector regresses and classifies the final 3D boxes. Thick black connections refer to the proposed single input AVOD architecture (SI-AVOD), while the thin blue connections refer to the baseline sensor fusion AVOD implementation (SF-AVOD).}
    \label{fig:avod}
\end{figure*}




In any case, object detection is inherently vulnerable to occlusions from external blockers, and the limited angular resolution of the sensors, especially in distant regions, may render the perception of the environment inaccurate.
In this context, more robust scene understanding could be achieved if vehicles shared their sensor data with cloud computing platforms (i.e., to leverage more computational resources to store and process data than available on board) as well as with other vehicles (i.e., to extend their perception range beyond the capabilities of their own instrumentation)~\cite{higuchi2019value}. 
Exchanging sensor data, however, can be challenging with standard communication technologies. For example, a raw LiDAR perception with 0.7 million points, generated  at 30~fps, would require a data rate of around 500 Mbps~\cite{cao20193d}: in comparison, the IEEE 802.11p standard for \gls{v2v} transmissions can only offer a nominal data rate of up to 27 Mbps~\cite{giordani2018feasibility}.

On one side, capacity issues can be partially solved by using the lower part of the \gls{mmwave} spectrum, as suggested by the latest IEEE and 3GPP specifications for future vehicular networks~\cite{zugno2020toward}.
A more efficient approach would be to carefully select the amount of data to be broadcast over bandwidth-limited channels~\cite{giordani2019investigating,giordani2019framework} to save network resources. 
The main concern lies in the choice of which sensor(s) would guarantee the optimal trade-off between channel occupancy and accuracy in the object detection task.

Notably, object detection methods are typically designed to accept single inputs only. For example, \gls{yolo}~\cite{redmon2016yolo} and Voxelnet~\cite{zhou2018voxelnet} have become very popular solutions to guarantee precise detections using camera's and LiDAR's inputs, respectively.
This implies, however, that several object detectors may need to be trained and executed at the same time, which may be difficult to~handle with limited computational resources.~Recently, sensor fusion algorithms like \gls{avod}~\cite{ku2018joint} were proven to ensure better reliability than single input schemes by exploiting together the complementary properties of cameras and LiDARs, even though they do~not permit detection when only one type of sensor data is accessible.


In this paper, we conduct experiments on the Kitti~\cite{geiger2012are} and Lyft~\cite{lyft2019} vision benchmark suites, and investigate whether sensor fusion can guarantee more accurate detections than single input methods, despite the additional traffic load on the channel for retrieving observations from multiple sensors. 
To do so, prior works typically considered separate sensor fusion or single input detection methods to be trained and executed in parallel on board of vehicles. For example, the authors in~\cite{xu2018pointfusion,li2019multi,vora2020pointpainting} proved the superior or competitive detection performance of sensor fusion schemes compared to conventional camera- and LiDAR-based approaches. 
On the other hand,  we addressed this problem by modifying the architecture of \gls{avod} to also accept single inputs from either cameras or LiDARs.
This allows our \gls{avod} implementation to be deployed as a standalone object detector, regardless of the amount (and class) of sensor data available, thus saving computational resources compared to state-of-the-art solutions.


We demonstrate that LiDAR only inputs can obtain an accuracy in the detection of large objects (like cars) similar to sensor fusion, which would require two different types of sensor data, thus mitigating the burden on the channel. On the contrary, sensor fusion can guarantee better detection for smaller road entities like pedestrians and cyclists by leveraging camera~images to make up for the sparsity of the LiDAR data.


The remainder of this paper is organized as follows.
In Sec. \ref{sec:avod} we describe the AVOD architecture and how it has been modified to accept single inputs. In Sec.~\ref{sec:simulation_setup} we present our simulation setup. In Sec.~\ref{sec:simulation_results} we illustrate our simulation results, while in Sec.~\ref{sec:conclusions_and_future_works} we summarize our conclusions and propose suggestions for future work.

\section{The AVOD Architecture for Object Detection} 
\label{sec:avod}

In the last few years, sensor fusion object detectors have appeared as a promising solution to generate robust detections by relying on different sensor modalities~\cite{chen2017multi}.
For example,  \gls{avod}~\cite{ku2018joint} fuses both front view RGB images and point clouds together, even though it does not permit detection when only one type of sensor data is available. 
In this section, we describe how we modified the baseline structure of sensor fusion \gls{avod} (SF-AVOD) to also accept, as an input, standalone camera images (SIC-AVOD) or standalone LiDAR point clouds (SIL-AVOD).



\subsection{Sensor Fusion AVOD (SF-AVOD)} 
\gls{avod} takes as an input two different types of data: a \gls{bev} projection, built from a LiDAR point cloud, and an image from an RGB camera. It then processes the inputs into two steps, namely the \emph{first stage detector} and \emph{second stage detector}, as illustrated in Fig.~\ref{fig:avod}.
\smallskip

\subsubsection{First stage detector}
It consists of a \gls{rpn} to regress non-oriented anchor proposals.
\smallskip

\begin{itemize}[leftmargin=*]
  \item \emph{Feature extractor.} The sensor inputs, after some preprocessing, are separately fed into two \glspl{cnn} to extract the feature maps. The feature extractor is inspired to the Feature Pyramid Network (FPN) in~\cite{lin2017feature} and has been structured according to the VGG16 model~\cite{simonyan2014very}: the resulting feature map was shown to represent both image and BEV inputs with 8 times lower resolution compared to the original data~\cite{ku2018joint}, which may limit the performance on smaller objects such as pedestrians and cyclists.
The output feature maps from both input streams are shared by the RPN in the first stage detector (in this case a $1\times1$ convolution is applied on the feature maps to reduce the dimensionality of subsequent crop-and-resize operations) and the second stage~detector.

\item \emph{3D anchor generation.} The system generates nine anchors for each feature point. They are predefined and non trainable, but their shape, number and size can be tuned manually. The choice of the shape can be based on either the feature map size or the size of the objects to be detected. 
In our case, AVOD is trained to operate in driving environments, where most objects are expected to be in a rectangular shape.
The size and placement of the anchors, instead, are determined by the characteristics of the training samples for each class of objects. 
After that, \gls{roi} pooling is applied to (i) extract feature crops for every anchor from each feature map, and (ii) resize the resulting feature crop to $3\times3$ to obtain equal-length feature vectors.

\item \emph{3D anchor top proposals.} The feature crops are fused together using an element-wise mean operation. 
The fused features are then randomly sub-sampled and fed into two task-specific branches consisting of fully connected layers: the first one defines whether an anchor box is ``background'' or ``object,'' and the second one produces regression offsets representing the difference between the ground truth's and the anchors' centroids and dimensions.
Anchor top proposals are selected by filtering redundant anchors via \gls{nms}, so that only those with an \gls{iou} larger than 0.7 with respect to ground truth boxes are maintained.
\end{itemize}	
\smallskip

\subsubsection{Second Stage Detector}
It is used to regress the final 3D boxes, and takes as inputs (i) the feature maps produced in the first stage detector and (ii) the anchor top proposals.
\begin{itemize}[leftmargin=*]

\item \emph{Bounding box generation.}  \gls{roi} pooling is applied to the feature maps from the first stage detector, which are cropped and resized for each anchor top proposal\footnote{Since in the second stage detector the number of top proposals is an order of magnitude lower than the number of anchors used in the first stage detector's RoI pooling, the feature maps are used without the $1\times1$~convolution.}, and passed into the detection network. 
Crops from both inputs are resized to $7\times7$, and fused with an element-wise mean~operation.
\smallskip

\item \emph{Bounding box classification.} The fused feature maps are fed into a single set of three fully connected layers to obtain regression, classification, and orientation of each box proposals.
Proposals with IoU less than 0.55 (0.45) for cars (pedestrians and cyclists) are classified as ``background.''
Finally, \gls{nms} is applied to remove overlapping detections, keeping the top bounding~boxes.
\end{itemize}	

\subsection{Single Input AVOD (SI-AVOD)} 
\label{sub:single_input_avod}

AVOD was designed to accept both LiDAR point clouds and RGB camera images as inputs. 
In this paper, the code has been altered to allow for accurate detection based on camera only inputs (SIC-AVOD) or LiDAR only input (SIL-AVOD), as displayed in Fig.~\ref{fig:avod} (thick black connections) and described~below.
\smallskip

\subsubsection{First Stage Detector}
The output of the $1\times1$ convolutional layer is fed into an \gls{roi} pooling layer, which is dimensioned freely to best represent the single input under consideration, i.e., unlike the SF-AVOD implementation, in this setup there is no need to match the features' size. 
Specifically, crop and resize parameters can now be configured in such a way that the detection performance of smaller objects (e.g., pedestrians and cyclists, which are generally penalized in the LiDAR input) improves.

The output of the RoI pooling is then directly fed into the first set of fully connected layers. In sensor fusion mode, dropout layers and batch normalization, as well as input masking, were added to each layer to scale the feature maps from both inputs, and to reduce overfitting, respectively. 
This was performed by randomly setting to zero proposal crops coming from either sample streams during training.
In single input mode, instead, the network has been simplified by removing the input masking layer, which was proven to degrade the detection performance significantly.

\subsubsection{Second Stage Detector}
The output of the first stage detector is fed into an RoI pooling layer. Simultaneously, features maps from the initial feature extractor are fetched, and fed also into an RoI pooling layer, whose parameters are chosen so that the outputs of the two RoIs match in size. 
Again, we disable input masking which, in sensor fusion, was added to randomly drop paths from one of the two sensor streams.

After that, feature crops from the second stage detector are processed by applying dropout and batch regolarizations, and directly sent into three separate sets of fully connected layers for classification, regression, and orientation of the bounding box proposals, respectively.





\section{Simulation Setup} 
\label{sec:simulation_setup}

\begin{table}[t!]
\vspace{0.33cm}
	\caption{Parameters of automotive sensors for Kitti and Lyft datasets.}
    \label{tab:sensors}
    \centering
\renewcommand{\arraystretch}{2}
\begin{tabular}{|l|l|c|c|}
\hline
\multicolumn{2}{|c|}{Parameters}                       & Kitti~\cite{geiger2012are} & Lyft~\cite{lyft2019} \\ \hline
\multirow{4}{*}{\rotatebox[origin=c]{90}{LiDAR}}  & Angular resolution & 0.08$^\circ$ azimuth      &   0.02$^\circ$ azimuth  \\ \cline{2-4} 
                        & Beams              &   64    &   64   \\ \cline{2-4} 
                        & Points/s           &    2.2 M   &  2.16 M    \\ \cline{2-4} 
                        & FPS                &    5--20 Hz   & 10 Hz     \\ \hline
\multirow{2}{*}{\rotatebox[origin=c]{90}{Camera}} & Pixel resolution   &   1384$\times$1032    & 1024$\times$2240     \\ \cline{2-4} 
                        & FPS                &    15 Hz   &    10 Hz  \\ \hline
\end{tabular}
\end{table}

In this section we present the datasets (Sec.~\ref{sub:dataset}), the training setup (Sec.~\ref{sub:training_setup}) and the performance metrics (Sec.~\ref{sub:performance_metrics}) that were used in our performance evaluation.

\subsection{Datasets} 
\label{sub:dataset}
The performance of the single input and sensor fusion AVOD architectures is evaluated on the three classes of objects (i.e., cars, pedestrians, and cyclists) of the Kitti~\cite{geiger2012are} and Lyft~\cite{lyft2019} multi-modal datasets.
Specifically, the Kitti sensor suite includes a Velodyne Laser scanner, two RGB cameras, and two greyscale cameras, mounted on a  Volkswagen Passat~\cite{geiger2012are}.
The Lyft setup consists of a center roof LiDAR, six wide-field-of-view cameras, and one front-view long focal length camera, mounted on a Renault Zoe \cite{lyft2019}.
More details about each sensor's specifications are reported in Table~\ref{tab:sensors}.

\begin{figure}[b!]
\centering
  \begin{subfigure}[b!]{0.44\textwidth}
  \centering
    \includegraphics[width=0.99\columnwidth]{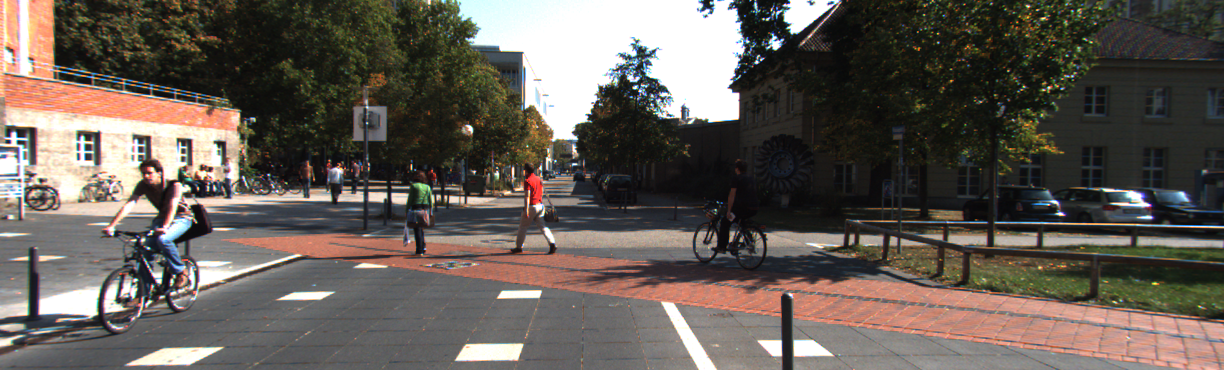}
    \caption{Front camera image from the Kitti dataset.}
    \label{fig:kitti}
  \end{subfigure} \\
  \begin{subfigure}[b!]{0.44\textwidth}
  \centering
    \includegraphics[width=0.99\columnwidth]{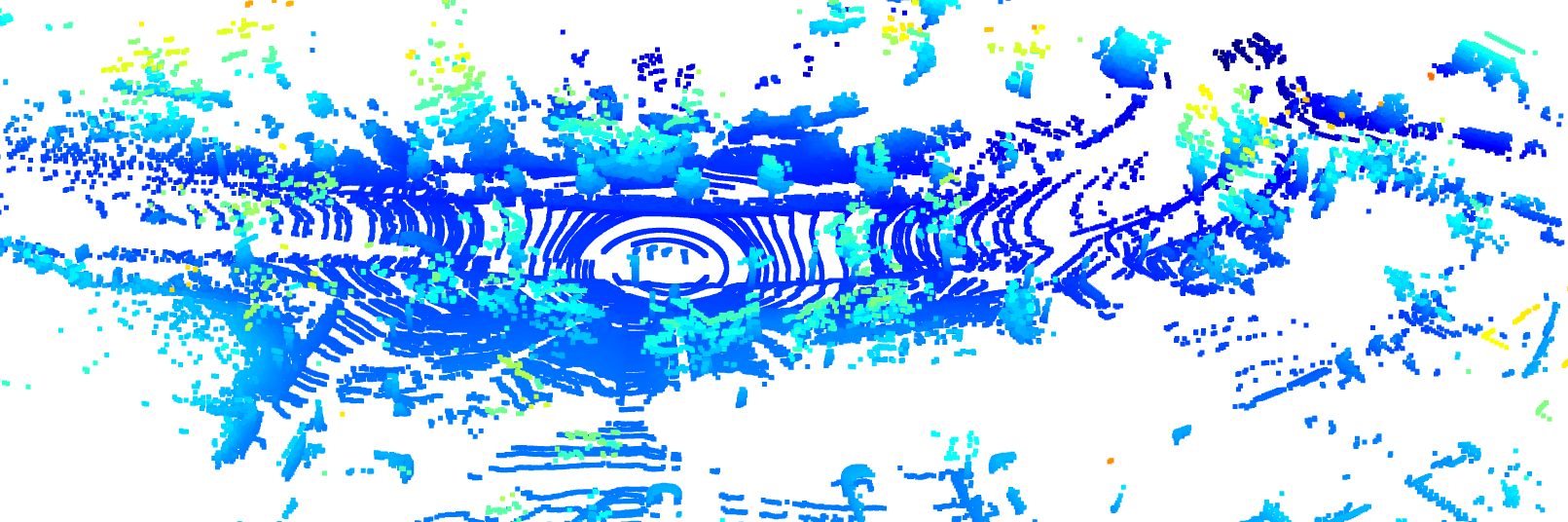}
    \caption{LiDAR point cloud from the Lyft dataset.}
    \label{fig:lyft}
  \end{subfigure}
  \setlength{\belowcaptionskip}{-0.33cm}
\caption{An example of scenes collected from the Kitti and Lyft datasets. }
	  \label{fig:capacity} 
\end{figure}



Besides the different sensor suites, the two datasets have different characteristics which make them equally valuable for this work.
First, the Kitti dataset contains several dedicated scenes where pedestrians and cyclists are clearly visible and recognizable, as illustrated in Fig.~\ref{fig:kitti}.
On the contrary, these classes of objects are extremely rare in the Lyft acquisitions, and appear either blurred, possibly for privacy reasons, or too far from the source sensor to be detected with sufficient accuracy.
The latter problem is particularly critical when considering LiDAR data, which are notoriously too sparse to efficiently capture fine-grained information from small objects over long distances~\cite{guo2020deep}, leading to false positives.
Second, unlike Kitti, the Lyft dataset incorporates scenes in different lighting and weather conditions, which may severely impact the detection performance.

	

\subsection{Training setup} 
\label{sub:training_setup}
For both Kitti and Lyft datasets, 7500 images were selected for training and 2500 for testing, randomly selected from the available scenes.
Given the inherently different characteristics of the two datasets, two separate networks were trained for the car and the cyclist/pedestrian classes, respectively.
The hardware we used was an HP Z1 Entry Tower, Intel core i7 vPro, equipped with an NVIDIA GeForce RTX 2080.
Convergence was achieved in 40 epochs using the Adam optimizer~\cite{frcnn_keras} and batch size 8, for a total runtime of about 8 hours.\\

\subsection{Performance Metric: the mAP} 
\label{sub:performance_metrics}
The performance of the proposed AVOD implementation for object detection has been evaluated under the \gls{map} metric over classes.
Specifically, the per-class \gls{ap} is related to the precision-recall curve~\cite{everingham2010pascal,flach2015precision}, where precision (recall) is defined as the percentage of correct predictions (correctly predicted objects) over all the predictions (objects).
Specifically, the precision-recall curve is obtained by sorting the predictions in such a way that those with highest precision and lowest recall are ordered first, and those with highest recall and lowest precision are ordered last.
Interpolating the curve as in~\cite{everingham2010pascal}, the \gls{ap} is thus computed as the area between the recall axis and the curve.
Finally, the \gls{map} is obtained by averaging the mean of the \glspl{ap} for all the considered classes of objects.

In this work we consider the AP$_{\rm BEV}$ score, which is the AP computed considering the projection of the bounding boxes on the \gls{bev} space.\footnote{In the remainder of this paper, for ease of notation and with no ambiguity, the AP$_{\rm BEV}$  and AP terms will be used interchangeably.} As typical in object detection tasks, only bounding boxes with an \gls{iou} greater than 0.5 are considered for the \gls{ap} computation.

\begin{figure}[t!]
    \centering
        \setlength\fwidth{0.9\columnwidth} 
    \setlength\fheight{0.5\columnwidth}
	\begin{tikzpicture}
	
\definecolor{color0}{RGB}{255,205,178}%
\definecolor{color1}{RGB}{255,180,162}%
\definecolor{color2}{RGB}{229,152,155}%
\definecolor{color3}{RGB}{181, 131, 141}%
\definecolor{color4}{RGB}{109, 104, 117}%

\definecolor{mycolor6}{RGB}{0, 109, 119}%
\definecolor{mycolor7}{RGB}{131, 197, 190}%
\definecolor{mycolor8}{RGB}{237, 246, 249}%
\definecolor{mycolor9}{RGB}{255, 221, 210}%
\definecolor{mycolor10}{RGB}{226, 149, 120}%
	\pgfplotsset{
tick label style={font=\scriptsize},
label style={font=\scriptsize},
legend  style={font=\scriptsize}
}

\begin{axis}[%
width=0.951\fwidth,
height=\fheight,
at={(0\fwidth,0\fheight)},
scale only axis,
				legend cell align={left},
				legend style={fill opacity=0.8, draw opacity=1, text opacity=1, draw=white!80!black, 	
		at={(1,1)},anchor=north east},
				tick align=outside,
				tick pos=left,
				x grid style={white!69.0196078431373!black},
				xmajorgrids,
				xmin=-0.4375, xmax=4.6875,
				xtick style={color=black},
				xtick={0,1.25,2.5,4},
				xticklabels={mAP,Car AP,Cyclist AP,Pedestrian AP},
				y grid style={white!69.0196078431373!black},
				ylabel={AP},
				ymajorgrids,
				ymin=0, ymax=0.85,
				ytick style={color=black},
			]
			
			\draw[draw=color0,fill=mycolor7, draw=black] (axis 	
				cs:-0.25,0) rectangle (axis cs:0,0.52);
			\draw[draw=color0,fill=mycolor7, draw=black] (axis 	
				cs:1,0) rectangle (axis cs:1.25,0.77);
			\draw[draw=color0,fill=mycolor7, draw=black] (axis 	
				cs:2.25,0) rectangle (axis cs:2.5,0.34);
			\draw[draw=color0,fill=mycolor7, draw=black] (axis 	
				cs:3.75,0) rectangle (axis cs:4,0.46);
			\addlegendimage{ybar,area legend,draw=color0,fill=mycolor7, draw=black};
			\addlegendentry{SF-AVOD}
		
			\draw[draw=color0,fill=mycolor9, draw=black,  postaction={pattern=north east lines, opacity=0.5}] (axis 
				cs:0,0) rectangle (axis cs:0.25,0.34);
			\draw[draw=color0,fill=mycolor9, draw=black,  postaction={pattern=north east lines, opacity=0.5}] (axis 
				cs:1.25,0) rectangle (axis cs:1.5,0.58);
			\draw[draw=color0,fill=mycolor9, draw=black,  postaction={pattern=north east lines, opacity=0.5}] (axis 
				cs:2.5,0) rectangle (axis cs:2.75,0.14);
			\draw[draw=color0,fill=mycolor9, draw=black,  postaction={pattern=north east lines, opacity=0.5}] (axis 
				cs:4,0) rectangle (axis cs:4.25,0.30);
			\addlegendimage{ybar,area legend,draw=color0,fill=mycolor9, draw=black,  postaction={pattern=north east lines, opacity=0.5}};
			\addlegendentry{SIC-AVOD}

			\draw[draw=color0,fill=mycolor10, draw=black,  postaction={pattern=dots, opacity=0.5}] (axis cs:
				0.25,0) rectangle (axis cs:0.5,0.49);
			\draw[draw=color0,fill=mycolor10, draw=black,  postaction={pattern=dots, opacity=0.5}] (axis cs:
				1.5,0) rectangle (axis cs:1.75,0.74);
			\draw[draw=color0,fill=mycolor10, draw=black,  postaction={pattern=dots, opacity=0.5}] (axis cs:
				2.75,0) rectangle (axis cs:3,0.29);
			\draw[draw=color0,fill=mycolor10, draw=black,  postaction={pattern=dots, opacity=0.5}] (axis cs:
				4.25,0) rectangle (axis cs:4.5,0.29);
			\addlegendimage{ybar,area legend,draw=color0,fill=mycolor10, draw=black,  postaction={pattern=dots, opacity=0.5}};
			\addlegendentry{SIL-AVOD}
			
		\end{axis}

\end{tikzpicture}
    \caption{Per-class AP and mAP scores with single input (camera or LiDAR only) and sensor fusion AVOD implementations, considering the Kitti dataset.}
    \label{fig:kitti_ap}
\end{figure}

\begin{figure}[t!]
    \centering
    \setlength\fwidth{0.9\columnwidth} 
    \setlength\fheight{0.5\columnwidth}
	\begin{tikzpicture}
	
\definecolor{color0}{RGB}{255,205,178}%
\definecolor{color1}{RGB}{255,180,162}%
\definecolor{color2}{RGB}{229,152,155}%
\definecolor{color3}{RGB}{181, 131, 141}%
\definecolor{mycolor5}{RGB}{109, 104, 117}%

\definecolor{mycolor6}{RGB}{0, 109, 119}%
\definecolor{mycolor7}{RGB}{131, 197, 190}%
\definecolor{mycolor8}{RGB}{237, 246, 249}%
\definecolor{mycolor9}{RGB}{255, 221, 210}%
\definecolor{mycolor10}{RGB}{226, 149, 120}%
	\pgfplotsset{
tick label style={font=\scriptsize},
label style={font=\scriptsize},
legend  style={font=\scriptsize}
}

\begin{axis}[%
width=0.951\fwidth,
height=\fheight,
at={(0\fwidth,0\fheight)},
scale only axis,
legend cell align={left},
legend style={fill opacity=0.8, draw opacity=1, text opacity=1, draw=white!80!black, 	
		at={(1,1)},anchor=north east},
tick align=outside,
tick pos=left,
x grid style={white!69.0196078431373!black},
xmajorgrids,
xmin=-0.4375, xmax=4.6875,
xtick style={color=black},
xtick={0,1.25,2.5,4},
xticklabels={mAP,Car AP,Cyclist AP,Pedestrian AP},
y grid style={white!69.0196078431373!black},
ylabel={AP},
ymajorgrids,
ymin=0, ymax=0.85,
ytick style={color=black},
			]
			\draw[draw=color0,fill=mycolor7, draw=black] (axis 	
				cs:-0.25,0) rectangle (axis cs:0.0,0.47);
			\draw[draw=color0,fill=mycolor7, draw=black] (axis 
				cs:1.0,0) rectangle (axis cs:1.25,0.72);
			\draw[draw=color0,fill=mycolor7, draw=black] (axis 
				cs:2.25,0) rectangle (axis cs:2.5,0.32);
			\draw[draw=color0,fill=mycolor7, draw=black] (axis 
				cs:3.75,0) rectangle (axis cs:4.0,0.38);
			\addlegendimage{ybar,area legend,draw=color0,fill=mycolor7, draw=black};
			\addlegendentry{SF-AVOD}
		
		
			\draw[draw=color0,fill=mycolor9, draw=black,  postaction={pattern=north east lines, opacity=0.5}] (axis 
				cs:0.0,0) rectangle (axis cs:0.25,0.18);
			\draw[draw=color0,fill=mycolor9, draw=black,  postaction={pattern=north east lines, opacity=0.5}] (axis 
				cs:1.25,0) rectangle (axis cs:1.5,0.39);
			\draw[draw=color0,fill=mycolor9, draw=black,  postaction={pattern=north east lines, opacity=0.5}] (axis 
				cs:2.5,0) rectangle (axis cs:2.75,0.1);
			\draw[draw=color0,fill=mycolor9, draw=black,  postaction={pattern=north east lines, opacity=0.5}] (axis 	
				cs:4.0,0) rectangle (axis cs:4.25,0.15);
			\addlegendimage{ybar,area legend,draw=color0,fill=mycolor9, draw=black,  postaction={pattern=north east lines, opacity=0.5}};
			\addlegendentry{SIC-AVOD}
		
		
			\draw[draw=color0,fill=mycolor10, draw=black,  postaction={pattern=dots, opacity=0.5}] (axis cs:
				0.25,0) rectangle (axis cs:0.5,0.39);
			\draw[draw=color0,fill=mycolor10, draw=black,  postaction={pattern=dots, opacity=0.5}]  (axis cs:
				1.5,0) rectangle (axis cs:1.75,0.71);
			\draw[draw=color0,fill=mycolor10, draw=black,  postaction={pattern=dots, opacity=0.5}]  (axis cs:	
				2.75,0) rectangle (axis cs:3.0,0.22);
			\draw[draw=color0,fill=mycolor10, draw=black,  postaction={pattern=dots, opacity=0.5}]  (axis cs:	
				4.25,0) rectangle (axis cs:4.5,0.25);
			\addlegendimage{ybar,area legend,draw=color0,fill=mycolor10, draw=black,  postaction={pattern=dots, opacity=0.5}};
			\addlegendentry{SIL-AVOD}
		
		\end{axis}
		
	\end{tikzpicture}
    \caption{Per-class AP and mAP scores with single input (camera or LiDAR only) and sensor fusion AVOD implementations, considering the Lyft dataset.}
    \label{fig:lyft_ap}
\end{figure}

\begin{figure*}[b!]
\centering
  \begin{subfigure}[b!]{0.32\textwidth}
  \centering
    \includegraphics[width=0.99\columnwidth]{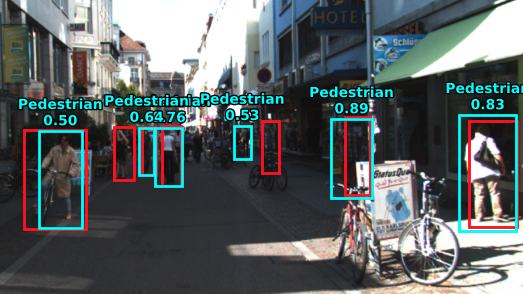}
    \caption{Object detection for SIC-AVOD.}
    \label{fig:cam-3d}
  \end{subfigure} \,
  \begin{subfigure}[b!]{0.32\textwidth}
  \centering
    \includegraphics[width=0.99\columnwidth]{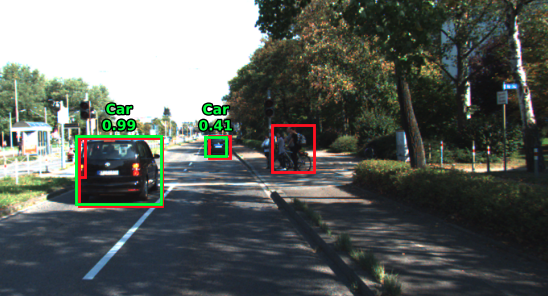}
    \caption{Object detection for SIL-AVOD (RGB projection).}
    \label{fig:lid-3d}
  \end{subfigure}\,
    \begin{subfigure}[b!]{0.32\textwidth}
  \centering
    \includegraphics[width=0.99\columnwidth]{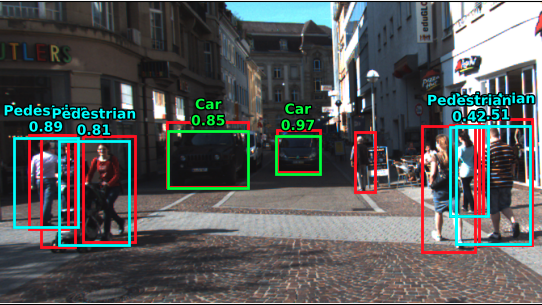}
    \caption{Object detection for SF-AVOD.}
    \label{fig:sf-3d}
  \end{subfigure}
  \setlength{\belowcaptionskip}{-0.33cm}
\caption{Qualitative comparisons of detection results for different AVOD implementations considering the Kitti dataset. The ground truth bounding boxes are depicted in red, while those estimated by SIC-AVOD, SIL-AVOD or SF-AVOD for the pedestrian/cyclist and car classes are in blue and green, respectively.}
	  \label{fig:3d} 
\end{figure*}

\section{Simulation Results} 
\label{sec:simulation_results}
In Sec.~\ref{ssec:si-sf} we compare through simulations the accuracy performance of single-input (SIC-AVOD and SIL-AVOD) vs. sensor-fusion (SF-AVOD) object detectors, which is assessed in terms of  \gls{map} as well as per-class AP.
In Sec.~\ref{ssec:accuracy}, the three schemes are compared also in terms of nominal data rate required to broadcast the sensor~data.


\subsection{Single Input vs. Sensor Fusion}
\label{ssec:si-sf}

\textbf{Dataset Comparison.}
Figs.~\ref{fig:kitti_ap} and \ref{fig:lyft_ap} show the \gls{ap} performance of single-input vs. sensor-fusion AVOD on the Kitti and Lyft datasets, respectively.
Notably, training on the Kitti data seems to offer a remarkable improvement in terms of mAP than Lyft's.
The difference is particularly significant when considering only the RGB input (in Lyft, the mAP for SIC-AVOD drops by almost 50\% compared to Kitti), whereas SIL-AVOD achieves good performance on both datasets (the mAP decreases by only 20\% when Lyft is considered).
This suggests that our implementation benefits from Kitti's horizontal (broad \gls{fov}) images rather than Lyft's squared (narrow \gls{fov}) images.
This is likely due to the fact that the former incorporate, on average, more informative data, as they capture wider road sections where cars, pedestrians and cyclists are located, while neglecting less significant sectors of the scene, e.g., the sky and the background.
Moreover, several Lyft scenes were acquired in rainy or foggy conditions, i.e., in low visibility for the cameras, or in busy roads where lined-up vehicles are difficult to distinguish for both sensors.
Specifically, different visibility conditions reduce the SIC-AVOD AP from 0.58 (Kitti) to 0.39 (Lyft) for the car class, and from 0.30 (Kitti) to 0.15 (Lyft) for the pedestrian class.
Finally,  it should be mentioned that the Lyft dataset contains a substantially lower number of pedestrian and cyclist instances than Kitti, thus exacerbating the well-know class imbalance problem~\cite{oksuz2019imbalance}, as discussed below.
\smallskip


\textbf{Class Comparison.}
As we already observed, the AP of the car class is much larger than those of the cyclist and pedestrian classes for both datasets.
On one side, smaller entities such as people far from the sensor are known to be more difficult to detect:
in RGB camera images, they may appear in few, noisy pixels, whereas in LiDAR point clouds they may be represented by very few points. 
At the same time, this difference in performance highlights the class imbalance problem~\cite{oksuz2019imbalance}, occurring when a dataset contains a disproportionate number of instances of one majority class (i.e., cars) with respect to other minority classes (i.e., cyclists and pedestrians).
As mentioned in Sec.~\ref{sec:simulation_setup}, we tried to mitigate this issue by training one network per class, even though the absence of pedestrians and cyclists in most scenes still heavily impacts the overall \gls{ap} performance.
\smallskip


\textbf{SIC/SIL/SF-AVOD Comparison.}
For this comparison, we adopt the Kitti dataset, so as to minimize the effects of class imbalance and image blurring.
Notably, AVOD was originally designed to jointly exploit 2D RGB camera images and 3D LiDAR point clouds: 
accordingly, Fig.~\ref{fig:kitti_ap} illustrates that SF-AVOD outperforms both SIC-AVOD and SIL-AVOD, achieving an overall \gls{map} of 0.52, while enhancing the \gls{ap} of the single classes.
The improvement is particularly clear for the detection of smaller objects, for which single-input AVOD struggles: for example, SF-AVOD improves the AP of the pedestrian class by 0.16 compared to SIC-AVOD.
The same conclusions can be drawn from Fig.~\ref{fig:3d}, which shows a qualitative example of object detection considering different AVOD architectures.
In particular, while SIL-AVOD (Fig.~\ref{fig:lid-3d}) strives to detect people and the farthest cars, which are represented by fewer points, SF-AVOD (Fig.~\ref{fig:sf-3d}) can well recognize different types of objects by also leveraging RGB images. 
However, it should be noticed that the mAP of SIL-AVOD decreases by only 6\% compared to SF-AVOD, in the face of a performance degradation by more than 35\% for SIC-AVOD. 
These results demonstrate that, when considering single inputs, the LiDAR data offer a significant advantage over the camera's, suggesting that SIL-AVOD may still represent an attractive option to perform robust and reliable detection, as discussed in the following section.


\begin{figure}[t!]
    \centering
        \setlength\fwidth{1.08\columnwidth} 
    \setlength\fheight{0.65\columnwidth}
    \setlength{\belowcaptionskip}{-0.33cm}
    \includegraphics[width=\columnwidth]{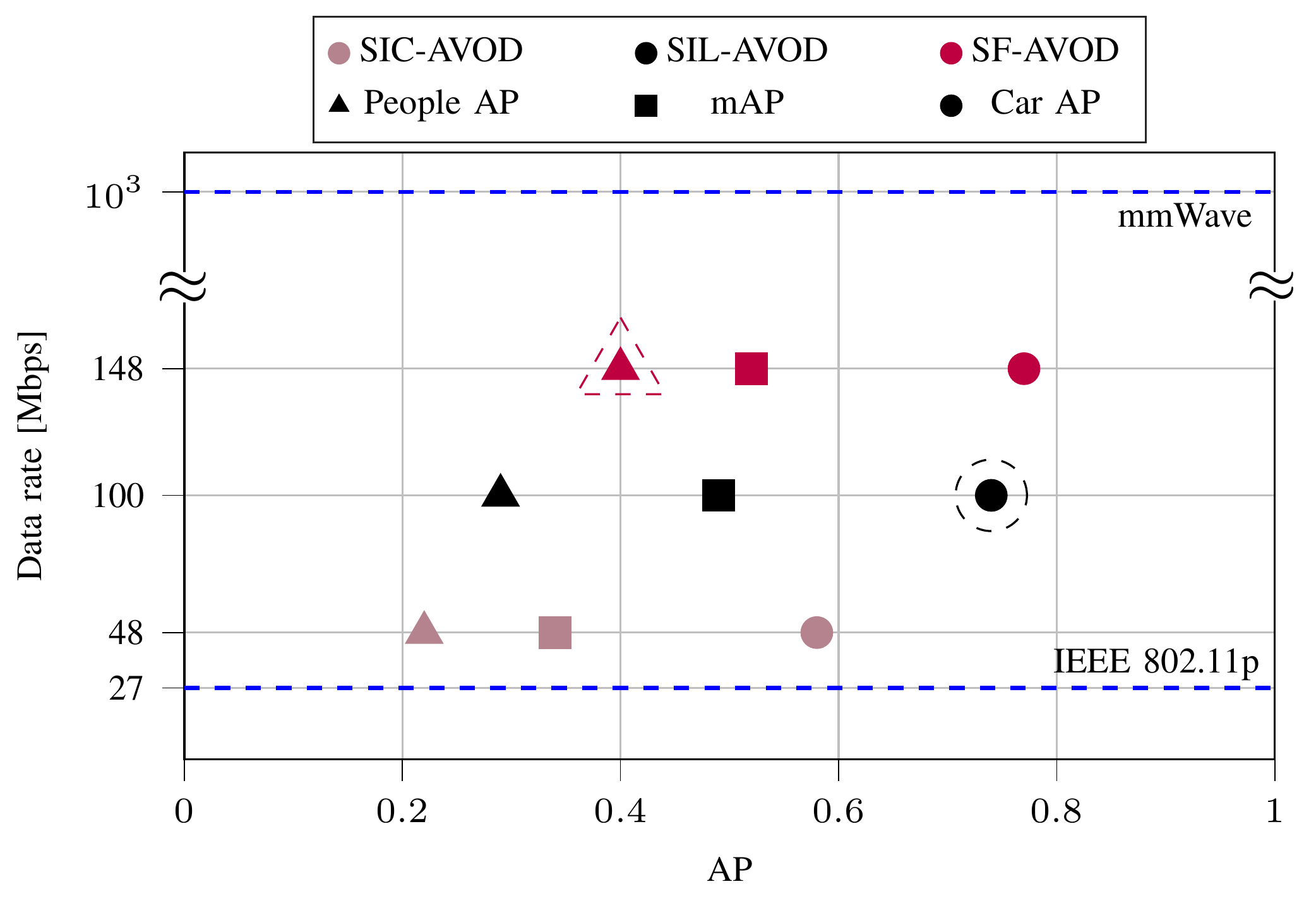}
    \caption{Data rate vs. detection accuracy, measured in terms of AP for different object classes, i.e., cars and people (which averages the AP of pedestrians and cyclists), considering SF-AVOD, SIC-AVOD, and SIL-AVOD  implementations in the Kitti dataset. The nominal capacity of IEEE 802.11p (\gls{mmwave}) technologies for V2V is set to 27 (1000) Mbps.}
    \label{fig:datarate_map}
\end{figure}

\subsection{Capacity vs. Accuracy Trade Off}
\label{ssec:accuracy}
Autonomous driving requires connected vehicles to share sensor data~\cite{arnold2020cooperative}, either to delegate object detection tasks to more powerful edge computing platforms, or to extend the perception range beyond line-of-sight.
In a bandwidth-constrained environment, where nodes compete for the limited channel resources, it is of foremost importance to select and transmit only the valuable data~\cite{higuchi2019value}.
In this section, moving from the \gls{ap} analysis, we provide guidelines towards the most appropriate object detection solution(s) to achieve high accuracy while minimizing communication costs.
More specifically, the sensing vehicles broadcast sensor data according to the type of input that each object detector requires: namely, either LiDAR or camera data for SIL-AVOD and SIC-AVOD, respectively, and a combination of the two for SF-AVOD.

The average data rates of the sensors are obtained from their data sheets.
Considering the Kitti suite, point clouds are produced from a 64-beam LiDAR in single mode at a rate equal to 12.48 Bytes/$\mu$s $\simeq $ 100 Mbps~\cite{velodyne}.
In turn, camera sensors generate data  at $1384\times1032\times24$ bit $\times\,10$ Hz $\simeq340$ Mbps, which can be reduced down to 12.24~Mbps assuming that PNG compression is applied~\cite{geiger2012are}.
Furthermore, whereas a LiDAR can offer a comprehensive view of the surroundings, the limited \gls{fov} of the cameras generally necessitates multiple sensors to cover the whole azimuth space. Thus, for a fair comparison, we consider a reference data rate of $12.24\times4=48.96$~Mbps for the RGB data, assuming that at least four cameras with a \gls{fov} of 90 degrees are employed.

Fig.~\ref{fig:datarate_map} shows the trade off between detection accuracy and data rate required to achieve it.
For the car class, SIL-AVOD achieves a 20\% AP gain over SIC-AVOD, while involving more than double its data rate (100 vs. 48 Mbps).
In turn, SF-AVOD would generate up to 148 Mbps to distribute data from both sensors, while increasing the AP by only 0.03.
Balancing the two, sending only the LiDAR data seems to be the most promising solution for detecting cars.
In support of this, connected vehicles may still exchange external data to improve localization (e.g., GPS coordinates), which makes it reasonable to trade a small accuracy drop for a $38\%$ saving on the channel occupancy.
On the contrary, non-connected, vulnerable road users such as pedestrians and cyclists should be ensured the greatest possible detection accuracy; thus, we identify SF-AVOD as the working point for these classes, achieving an average AP equal to 0.4 (vs. 0.3 of SIL-AVOD and 0.22 of SIC-AVOD) at $148$~Mbps.

As a concluding remark, Fig.~\ref{fig:datarate_map} shows the nominal capacity provided by some transmission technologies currently considered for vehicular communications.
Specifically, although advanced data compression techniques may lift the burden on the channel, the $27$~Mbps offered by the IEEE 802.11p standard cannot sustain the LiDAR data flow.
In turn, at most two simultaneous camera data streams can be transmitted using PNG compression: in this case, RGB images can still provide a means of detecting critical road entities such as pedestrians, though with low accuracy, as illustrated in Fig.~\ref{fig:cam-3d}.
Conversely, \glspl{mmwave} and other high-frequency communication technologies can offer multi-Gbps data rates, thus satisfying network requirements~\cite{zugno2020toward}.

\section{Conclusions and Future Works} 
\label{sec:conclusions_and_future_works}
Beyond-line-of-sight sensing is possible if connected vehicles share perception data with each other. In this work we analyzed how the number and type of sensors affect the object detection performance in real automotive scenarios, and studied how this impacts the channel occupancy. 
To do so, we altered the code of AVOD, a state-of-the-art object detector, to accept as inputs camera images, LiDAR point clouds, or a combination of the two. Then, we demonstrated that, while LiDAR only perceptions can provide high accuracy for detecting large objects such as cars, a sensor fusion approach would guarantee more robust detections for vulnerable road users such as pedestrians, despite consuming more channel~resources.

As part of our future work, we will investigate whether and how detection accuracy would improve using more than two input sensors, and analyze the trade-off between distributed and centralized object detection. Furthermore, a more accurate (system-level) simulation of the network will be considered.

\bibliographystyle{IEEEtran}
\bibliography{bibliography.bib}

\end{document}